\documentclass{aa}
\usepackage{xcolor}
\usepackage{graphicx}
\usepackage{txfonts}
\usepackage[colorlinks]{hyperref}
\usepackage{ulem}
\begin{document} 

\newcommand{\rga}[1]{\textcolor{blue}{#1}}
\newcommand{\ch}[1]{\textcolor{purple}{#1}}

   \title{Non-common path aberration compensation and a dark hole loop with a pyramid adaptive optics system: Application to SAXO+}

   \author{C. Goulas \inst{1}\fnmsep\thanks{\email{raphael.galicher@obspm.fr}}
          \and R. Galicher \inst{1}
          \and F. Vidal \inst{1}
          \and J. Mazoyer \inst{1}
          \and F. Ferreira \inst{1}
          \and A. Sevin \inst{1}
          \and A. Potier \inst{1}
          \and A. Boccaletti \inst{1}
          \and E. Gendron \inst{1}
          \and C. Béchet \inst{2}
          \and M. Tallon \inst{2}
          \and M. Langlois \inst{2}
          \and C. Kulcsár \inst{3}
          \and H-F. Raynaud \inst{3}
          \and N. Galland \inst{3}
          \and L. Schreiber \inst{4, 7}
          \and I. Bernardino Dinis \inst{5}
          \and F. Wildi \inst{5}
          \and G. Chauvin \inst{6}
          }

   \institute{LIRA, Observatoire de Paris, Université PSL, Université Paris Cité, Sorbonne Université, CNRS, 5 place Jules Janssen, 92195 Meudon, France
             \and
             CRAL, CNRS, Université Claude Bernard Lyon 1, ENS de Lyon
             \and
             IOGS, CNRS, Laboratoire Charles Fabry, Université Paris-Saclay
             \and
             INAF, Osservatorio di Astrofisica e Scienza dello Spazio di Bologna
             \and
             Dept. of Astronomy, University of Geneva
             \and
             Laboratoire J.-L. Lagrange, CNRS, OCA, Université Côte d’Azur
             \and
             IPAG, CNRS, Université Grenoble Alpes}

   \date{Received September 15, 1996; accepted March 16, 1997}

 
  \abstract
   {In ground-based high-contrast instruments, residual aberrations known as non-common path aberrations (NCPAs) limit detection performance, as they are unseen by the adaptive optics (AO) wavefront sensor but impact the astrophysical image, creating quasi-static speckles. SAXO+, the upgrade of the SAXO (SPHERE AO system) includes a second loop of AO downstream of the SAXO loop. This second loop is equipped with a near-infrared pyramid wavefront sensor whose nonlinearities, usually described with modal optical gains, might be challenging for removing quasi-static speckles.}
   {We investigated two methods of quasi-static speckle removal : NCPA compensation and a dark hole loop, behind a pyramid AO system, measuring the interest of compensating for the pyramid optical gains.}
   {We performed end-to-end numerical simulations of SAXO+ under various astrophysical conditions (seeing, star magnitude). We offset the pyramid wavefront sensor operating point to apply both the speckle suppression methods, calibrating or not calibrating the optical gains. We evaluated their performance by measuring the residual starlight in the coronagraph image.}
   {A by-product of our study is an on-sky calibration method of measuring the pyramid optical gains. Non-common path aberration compensation reduces the residual starlight in the coronagraph image by a factor of up to 20 for seeing between 0.7" and 1" for a bright star and a factor of 2 at 0.7" for a faint star. Optical gains compensation enhances the performance at poor seeing and small pyramid modulation radius with a bright star, but shows a useless or even negative impact due to estimation inaccuracies at faint targets. On the other hand, the dark hole loop reduces the residual starlight by a factor of up to 200. The optical gain calibration enhances the dark hole performance behind a single pyramid AO system but is useless behind the SAXO+ system. Our parametric study gives baseline values for the efficient control of the dark hole loop for the SAXO+ system.}
   {}


   \maketitle

\section{Introduction}
Direct detection of exoplanet light paves the way for spectroscopy of exoplanet atmospheres. The main issues are the on-sky angular proximity and the flux ratio between the exoplanet and its host star. Thanks to a coronagraph, an exoplanet imaging instrument cancels the light from the star in the final image while maximizing the exoplanet image throughput. As current coronagraphs are designed to work with a perfectly flat wavefront, their performance is strongly degraded by the Earth-atmosphere turbulence and manufacturing defects in the optics of the instrument. The former are compensated for by an adaptive optics (AO) system. The latter create static and quasi-static aberrations. Some of these cannot be compensated for by the AO system because the coronagraph does not see the same aberrations as the AO wavefront sensor (WFS). Therefore, the AO system does not correct exactly what the astrophysical path sees. These are the non-common path aberrations (NCPAs) between the wavefront sensing path and the astrophysical path. The NCPAs induce a speckle pattern in the coronagraph image that slowly evolves over time because of thermal variations and the flexure of mechanical structures, among other things, and mimics the signal of an exoplanet. If the AO system sufficiently compensates for the atmospheric turbulence, the coronagraph image is dominated by the NCPA speckles. Several strategies have been proposed to counterbalance the NCPA effects, as is described in section~\ref{sec:ncpa_comp}.

Multiple instruments combining AO systems and coronagraphs have been installed since 2010  : GPI at Gemini South \citep{Macintosh_2018}, Clio2/MagAO at the \textit{Magellan} telescope \citep{Sivanandam_2006, Close_2010}, SCExAO at Subaru \citep{Jovanovic_2015}, and SPHERE at the VLT \citep{Beuzit_2019}. The SPHERE AO system is called SAXO \citep{Fusco_2014} and includes a Shack-Hartmann~\citep[SH,][]{Shack_1971} WFS working in visible light, whereas the coronagraph images are recorded in the near-infrared. Thanks to SAXO, SPHERE can detect young Jupiters $10^5$ fainter than their host star at a few hundred milliarcseconds (mas) from the star. In good conditions and for bright targets, the coronagraph images are dominated by stellar speckles, and the minimization of speckle intensity with a so-called dark hole technique has been demonstrated using the SAXO deformable mirror~\citep[DM,][]{Potier_2022,Galicher_2024}. However, most of the time the AO system is limited by the WFS sensitivity and the temporal error of the loop \citep{Cantalloube_2020}. To address these issues and improve the detection capabilities, an upgrade of SAXO, called SAXO+, is currently in the design phase \citep{Boccaletti_2024}. GPI and MagAO are also being upgraded to GPI 2.0 \citep{Chilcote_2022} and MagAO-X \citep{Males_2022}, respectively.

\begin{figure}
    \centering
    \includegraphics[width=0.9\hsize]{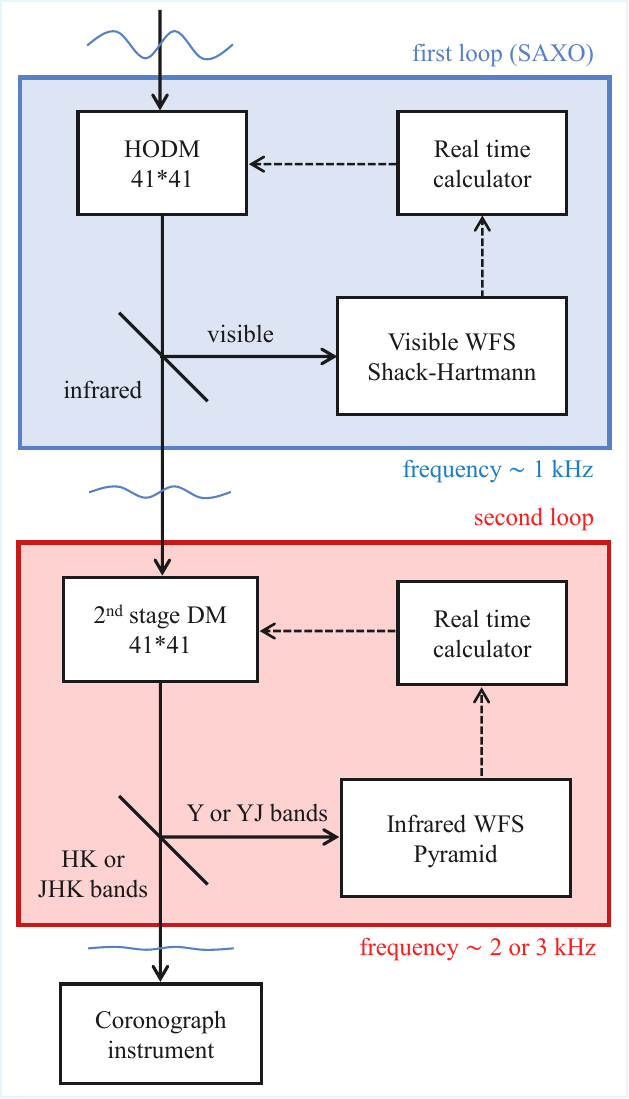}
    \caption{SAXO+ current design. The first stage maximum frequency is 1.38 kHz and second stage maximum frequency is 3 kHz.}
    \label{fig:design}
\end{figure}

The SAXO+ objective is to provide a wavefront stable enough that the coronagraph images are speckle-dominated most of the time. The SAXO+ design includes a second AO loop, downstream of the current SAXO stage, as is described in Fig. \ref{fig:design}. The first stage is a 1.38 kHz AO loop with a 40 $\times$ 40 SH WFS in visible light and a 41 $\times$ 41 DM. In SAXO+, the second stage is faster, at a maximum speed of 3 kHz, to address the temporal error of SAXO. The wavefront sensing is done with near-infrared light, at 1.2 \textmu m, with a 50 $\times$ 50 pyramid WFS \citep[PWFS,][]{Ragazzoni_1996}, more sensitive than the SH WFS. The second loop DM is a $28\times28$ kilo-DM from Boston Micromachines. As the actuators on the edge of the DM are not fully controllable, we use only 26 actuators across the telescope pupil diameter. There are two distinct real-time computers, one for each loop. However, the control algorithms of the second stage account for the closed-loop telemetry of the first stage, although in this paper we assume two independent loops. Such cascade AO systems have shown encouraging results in simulations by \citet{Cerpa-Urra_2022} and experimentally on an optical testbed by \citet{NDiaye_2023}.

To minimize the intensity of the quasi-static speckle during a closed-loop run with SAXO, tests took place in which the reference slopes of the SH WFS were offset so that the DM corrects for the NCPAs seen by the coronagraph \citep{Vigan_2019, Potier_2022}. In SAXO+, the correction cannot be done by applying an offset on the first stage only anymore, because the second loop would see this correction and react backward, canceling the first stage offset. Thus, it must be performed through the second stage, offsetting the reference measurement of the pyramid WFS.

In this paper, we study how to offset a pyramid AO loop to eliminate the quasi-static speckles. We assume two AO systems: a single-stage pyramid AO system and the SAXO+ system. In Section~\ref{sec:ncpa_comp}, we define two speckle removal techniques, a calibration that is done only once before the observation (NCPA compensation) and a dark hole loop that updates the speckle minimization during the astrophysical observation. In Section~\ref{sec:sim_assumptions}, we present the numerical simulation tool and assumptions. In Section~\ref{sec:opti_gains}, we propose a calibration strategy to measure the PWFS optical gains. In Section~\ref{sec:ncpa}, we study the interest in compensating the PWFS optical gains for the NCPA compensation. In Section~\ref{sec:dh}, we present a parametric study of the active NCPA compensation using the dark hole technique (star magnitude, turbulence, with and without PWFS optical gains, dark hole loop gain).

\section{Speckle removal techniques}
\label{sec:ncpa_comp}

\subsection{NCPA compensation or dark hole loop}

There are several ways to calibrate and correct for quasi-static speckles in the coronagraph image, which evolve slowly with respect to the exposure time of the image. We can estimate the NCPAs once and then offset the AO loop to minimize the phase in the pupil plane before the coronagraph. This phase minimization technique, sometimes known as phase conjugation, is what we call the NCPA compensation in this paper. In SAXO, \citet{Vigan_2019} measured the NCPAs using a Zernike WFS and then offset the SH WFS reference slopes to compensate for the NCPAs in the pupil plane upstream from the coronagraph. They repeated this procedure for a few iterations by closing a feedback loop, while the AO loop was closed. But in our study, we assumed that we knew perfectly the NCPAs introduced in the coronagraph channel, and updated the offset of the pyramid reference only once before recording the coronagraph image. The NCPA compensation minimizes the phase aberrations in the pupil plane, decreasing the stellar speckle intensity in the coronagraph image. However, such NCPA compensation does not minimize the starlight intensity due to amplitude aberrations or diffraction effects of the spiders.

A dark hole loop suppresses such starlight residuals in the coronagraph image, as the dark hole algorithm aims to minimize the quasi-static intensity in a given area of the focal plane. For example, in SPHERE, \citealt{Potier_2022} estimated the speckle electric field in the coronagraph image using the Pair-Wise Probing (PWP) technique \citep{Give'on_2007}. The PWP records four coronagraph images poking two actuators of the DM with a known amplitude so that the speckle intensity is time-modulated. Once the electric field was known, they used the electric field conjugation \citep[EFC,][]{Give'on_2007} method to minimize the starlight intensity in a given area of the coronagraph image by optimizing the DM shape. Usually, the dark hole loop repeats several iterations of sensing and minimizing the electric field because of linear assumptions or errors of the light-propagation model. The DH loop minimizes the speckle intensity whatever their origin, phase or amplitude aberrations. Amplitude aberrations correspond to any throughput variations inside the circular pupil, such as diffraction due to telescope spiders and central obstruction (or small local variations in the instrument transmission, but we did not simulate it). In the SAXO+ system, the dark hole loop will be the third loop of correction in a cascade of the two existing AO loops.

The NCPA compensation and dark hole loop performance behind a pyramid WFS are presented in sections~\ref{sec:ncpa_comp} and~\ref{sec:dh}. One strategy would be to combine both techniques, but in this paper, we consider them separately.

\subsection{The pyramid optical gains issue}

As was already mentioned, to modify the DM shape for NCPA sensing (actuator poke for PWP) or control (change the DM shape for EFC correction or NCPA compensation) while the AO loop is running, one needs to offset the PWFS whose difficulty depends on the actual wavefront regime. Usually, the PWFS response to the DM modal basis is calibrated in a linear, diffraction-limited regime and stored in the interaction matrix. But during an observation, the PWFS response differs from the calibration regime. The correction residuals of the AO loop yield a sensitivity loss for the PWFS compared to the calibrated interaction matrix. This nonlinearity can be described with modal coefficients called optical gains \citep{Deo_2019, Chambouleyron_2021}. The optical gains depend on the spatial frequency of the mode and decrease as the seeing increases. As the optical gains quantify a sensitivity loss on sky compared to the interaction matrix, they are between 0 and 1. If they are significantly below 1, the optical gains might be an issue for the SAXO+ AO performance itself (see \citealt{Goulas_2024}) and an issue when correcting the NCPAs or running a dark hole loop.

We call $\phi$ the phase to apply to the DM (to compensate for the NCPAs or to apply a dark hole specific shape), and $g$ the optical gain associated with this phase for a given observing condition. Offsetting the PWFS from the linear, diffraction-limited interaction matrix, the AO loop will apply $\phi/g$ on the~DM resulting in a time-dependant overshoot. Similarly, if the optical gain of the actuator poking is too small, the poke amplitude used for PWP will be larger than expected, biasing the focal plane wavefront sensing. Moreover, we can wonder if the system will correctly poke a single actuator as the pyramid AO loop uses a modal basis. All these questions are addressed in Sections~\ref{sec:ncpa} and~\ref{sec:dh}.

In the rest of the paper, we call the quantity existing in the PWFS measurement space, used to offset the reference measurement of the PWFS, the "PWFS reference offset". This offset adds a static phase to the~DM that is seen in the astrophysical channel (here the coronagraph channel). We explain here how we modified the PWFS reference offset when we took into account the optical gains. The offset can be the PWFS measurements, $\vec{m}_1$, that the phase, $\phi$, would give in a linear and diffraction-limited regime. To account for the PWFS optical gains, the PWFS reference offset, $\vec{m}_2$, was obtained from~$\vec{m}_1$ by applying the optical gain in the modal basis of the AO system:
\begin{equation}
    \vec{m}_2 = \vec{I} \cdot \vec{G} \cdot \vec{C} \cdot \vec{m}_1
\end{equation}
where
\begin{itemize}
    \item $\vec{m}$ is the measurement vector of the PWFS containing the $n_{pixel}$ values of the valid pixels on the PWFS camera (meaning, the pixels that receive flux when there is no aberrations),
    \item $\vec{I}$ is the modal interaction matrix, with $n_{pixel}$ rows and $n_{modes}$ columns,
    \item $\vec{C}$ is the modal command matrix (pseudoinverse of $\vec{I}$), with $n_{modes}$ rows and $n_{pixel}$ columns, and
    \item $\vec{G} = (g_i)$ is a diagonal matrix that contains the calibrated optical gains.
\end{itemize}

\section{Simulation assumptions}
\label{sec:sim_assumptions}

The numerical simulations were performed with COMPASS (fifth version), an end-to-end AO simulation tool \citep{Gratadour_2016}. COMPASS simulates atmospheric turbulence, telescope pupil, AO subsystems (SH and pyramid WFS, DM, control algorithm) and image formation. The models of each of these components are described on the official website \footnote{\url{https://compass.pages.obspm.fr/website/}}. Most of the simulation assumptions are described in \citet{Goulas_2024}. However, in the current paper, we did not use the CLOSE algorithm \citep{Deo_2021} to address the PWFS optical gain issue. We calibrated the modal optical gains thanks to the \citealt{Esposito_2020} method, discussed in Section~\ref{sec:opti_gains}.

The simulation parameters are presented in Table \ref{tab:param_1}. Depending on the section, the explored parameter space includes seeings, the pyramid modulation radius, and two science cases. These science cases are called « bright » and « faint » hereafter and are detailed in Table \ref{tab:param_2}. We chose a bright case to assess the best achievable performance by removing quasi-static speckles and compensating for optical gains, and a fainter and redder case as it is one of the main science cases for the SAXO+ upgrade. By default, the pyramid modulation radius is set to 3 $\lambda_{\mathrm{WFS}} / D$ if its value is not mentioned.

\begin{table}
    \renewcommand{\arraystretch}{1.12}
    \centering
    \caption{Simulation parameters. The asterisk * indicates the variable parameters of this study. The other ones are fixed parameters}
    \begin{tabular}{ll}
    \hline
        \multicolumn{2}{c}{\textbf{Turbulence}} \\
    \hline
        profile & ESO 35 layer median \\
        outer scale & $L_0$ = 25 m \\
        seeing* & $s$ = 0.4, 0.7, 1, 1.5, 2 arcsec \\
        coherence time & $\tau_0$ = 2 or 3 ms \\
    \hline
        \multicolumn{2}{c}{\textbf{Telescope}} \\
    \hline
        diameter & $D = 8$ m \\
        entrance pupil & VLT pupil with spiders \\
    \hline
        \multicolumn{2}{c}{\textbf{First stage}} \\
    \hline
        WFS type & 40 $\times$ 40 SH \\
        wavelength & $\lambda_{\mathrm{WFS}}$ = 700 nm \\
        readout noise & 0.1 electrons/pixel \\
        DM geometry & 41 $\times$ 41 + tip-tilt mirror \\
        loop delay & 0.84 ms \\
    \hline
        \multicolumn{2}{c}{\textbf{Second stage}} \\
    \hline
        WFS type & 50 $\times$ 50 pyramid \\
        wavelength & $\lambda_{\mathrm{WFS}}$ = 1.05 or 1.2 \textmu m \\
        readout noise & 0.24 electrons/pixel \\
        modulation radius* & 3, 2 and 1 $\lambda_{\mathrm{WFS}} / D$ \\
        DM geometry & 26 $\times$ 26 \\
        loop delay & 0.33 ms \\
    \hline
        \multicolumn{2}{c}{\textbf{Image formation}} \\
    \hline
        wavelength & $\lambda$ = 1.65 \textmu m \\
        NCPA amplitude & 45 nm RMS \\
    \end{tabular}
    \label{tab:param_1}
\end{table}

\begin{table}[]
    \renewcommand{\arraystretch}{1.12}
    \centering
    \caption{Science cases and associated parameters}
    \begin{tabular}{llll}
        \hline
        Science case & Bright & Faint \\
        \hline
        G-mag & 5.5 & 11.9 \\
        J-mag & 5.2 & 8.5 \\
        SH flux [ph-e/subap/ms] & 370 & 1.0 \\
        Pyr flux [ph-e/subap/ms] & 27 & 2.2 \\
        Pyr wavelength $\lambda_{\mathrm{WFS}}$ & 1.05 \textmu m & 1.2 \textmu m \\
        1st stage frequency & 1000 Hz & 300 Hz \\
        2nd stage frequency & 3000 Hz & 3000 Hz \\
        1st stage gain & 0.4 & 0.05 \\
        2nd stage gain & 0.5 & 0.3 \\
    \end{tabular}
    \label{tab:param_2}
\end{table}

For all of the coronagraph images in this paper, we considered neither photon noise nor detector noise, and the signal is monochromatic, at 1.65 \textmu m. The images were normalized to the maximum of the point spread function (PSF). The PSF was computed the same way as the image but without the coronagraph focal plane mask in the optical path. Hence, the normalized intensity in the images (usually known as the term raw contrast in coronagraphy) directly gives the performance in the planet/star luminosity ratio. In Section \ref{sec:ncpa}, we assume a perfect coronagraph~\citep{Cavarroc_2006} to focus on the NCPA compensation performance. For the dark hole simulations in Section~\ref{sec:dh}, we used the APLC coronagraph employed in SPHERE \citep{Carbillet_2011}. However, the simulated version does not feature the dead actuator patches of SPHERE.

The SPHERE NCPA were measured by Zelda~\citep{Vigan_2022}. \citet{Mazoyer_2024} extracted the power spectrum density (PSD) from the reconstructed NCPA phase screens, assuming an azimuthal invariant PSD, and cropping it beyond $20\,\lambda / D$. In this paper, we used NCPAs generated from this PSD and scaled to~$45\,$nm RMS, which is the expected amplitude for SAXO+. We only introduced phase aberrations in the pupil but no amplitude aberrations.


\section{Pyramid optical gains estimation}
\label{sec:opti_gains}
\subsection{Overview of the method}

To calibrate the PWFS modal optical gains, we used the method developed by \citet{Esposito_2020}. They simulated and tested it on sky, successfully compensating for the NCPAs (astigmatism) on the LBT telescope with the FLAO system \citep{Esposito_2010} and the LUCI2 instrument \citep{Heidt_2018}. The method consists in temporally modulating one mode of the modal basis while the AO loop is closed. Then, the optical gain of this mode was computed from the telemetry data (pyramid measurement and DM commands). \citet{Esposito_2020} measured the optical gain for only one mode, the astigmatism. In this paper, we propose a method to enable the calibration of the optical gain for each mode of the AO basis. Moreover, we used the optical gain as is explained by~Eq.\,\ref{eq:offset_with_bias}, whereas \citet{Esposito_2020} modified the control law.

We shall now describe our method based on the notation used in \citet{Esposito_2020}. We call~$\vec{c}$ the command vector containing the $n_{actus}$ actuator voltages computed by the integrator, and $\vec{B} = (\vec{b}_i)$ the modal basis matrix containing the $n_{modes}$ controlled modes, with $n_{actus}$ rows and $n_{modes}$ columns. To measure the optical gain $g_i$ related to the mode $\vec{b}_i$ in closed loop, we added a sinusoidal modulation of the mode~$\vec{b}_i$ to the command~$\vec{c}$. The voltages, $\vec{v}(t)$, applied to the DM read
\begin{equation}
    \vec{v}(t) = \vec{c}(t) + a \sin(2 \pi f t)\, \vec{b}_i,
    \label{eq:offset_with_bias}
\end{equation}
with two adjustable parameters, $a$ the modulation amplitude expressed in nanometer RMS across the pupil and $f$ the modulation frequency. The third parameter is the exposure time of the modulation, called $t_{modu}$. During this exposure time, we recorded both the command, $\vec{c}(t)$, and the pyramid measurement, $\vec{m}(t)$. Their projections on the mode~$\vec{b}_i$ are, respectively, the coordinate number~$i$ of the vectors $^\mathrm{t}\vec{B} \cdot \vec{c}(t)$ and $\vec{C} \cdot \vec{m}(t)$ with $\cdot$ being the matrix product. We demodulated these two signals to compute the optical gains related to the mode~$\vec{b}_i$ as the ratio of the command that is really sent to the DM and the command that is measured by the PWFS:
\begin{equation}
\label{eq:og}
    g_i = \frac{D([^\mathrm{t}\vec{B} \cdot \vec{c}(t)]_i)}{D([\vec{C} \cdot \vec{m}(t)]_i)},
\end{equation}
where $[\vec{x}]_i$ means the coordinate~$i$ of the vector~$\vec{x}$ and~$D$ is a demodulation operator. For a signal~$x(t)$, the demodulation is defined as
\begin{equation}
    D(x) = \frac{2}{t_{modu}} \left| \int_0^{t_{modu}} x(t)\exp(2 j \pi ft) dt \right|
\end{equation}

\begin{figure*}
    \centering
    \includegraphics[width=0.9\linewidth]{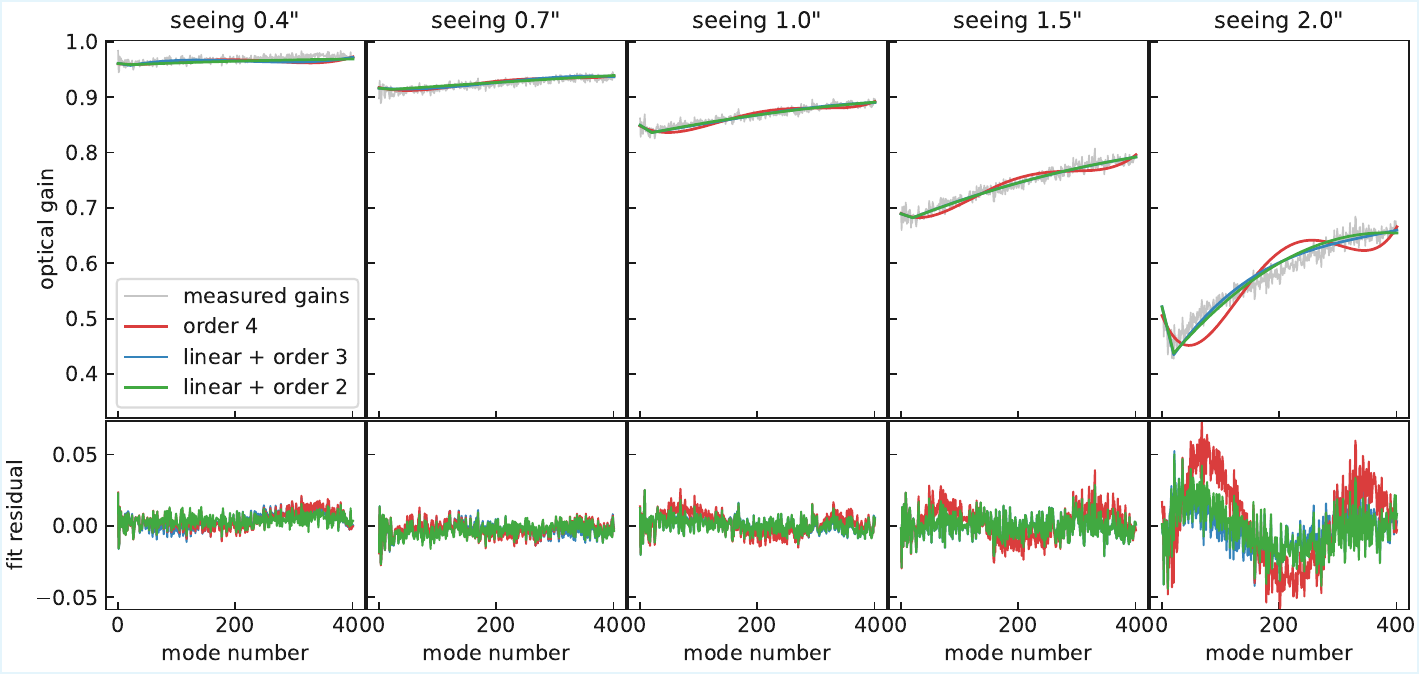}
    \caption{Optical gain versus the mode order, for several seeings. Top : Modal gains measurement (gray) and fit functions (red, blue and green). Bottom : Difference between the fit and the measurement. Science case : Bright,  $\tau_0 = 2$ ms.}
    \label{fig:fit_og}
\end{figure*}

The optical gains are strongly dependent on the observation conditions (seeing, coherence time, etc). Therefore, they need to be measured as soon as the conditions change, meaning on the scale of a minute or a couple of minutes. And measuring the optical gain of hundreds of modes as shown in gray in Figure~\ref{fig:fit_og} would last about~$1\,$min. Hence, the measured gains would never be up to date and a continuous calibration would be needed. Moreover, modulating the modes would add wavefront aberrations in the coronagraph channel degrading the coronagraph performance. To overcome these two limitations, we propose to speed up the optical gain calibration measuring the gains for 12 modes. This takes~$1$ to~$2\,$s. Hence, it can be done during the astrophysical detector overhead and it enables a fast update of the PWFS optical gain to adapt to the changing conditions. Once the~12 mode gains are measured, we adjust a synthetic function to determine the optical gain of all modes of the AO loop. Figure~\ref{fig:fit_og} also plots three polynomial fit curves using the~12 measured gains:
\begin{itemize}
    \item a fourth-order polynomial in red,
    \item a linear function for mode order below 20 and a third-order polynomial above 20 in blue,
    \item a linear function for mode order below 20 and a second-order polynomial above 20 in green.
\end{itemize}
The modulation frequency, $f$, is 200 Hz, the modulation amplitude~$a = 10\,$nm and the exposure time~$t_{modu} = 0.5\,$s. The simulation was done with the SAXO+ system, and we modulated the second stage~DM. The transition at mode order~$20$ in the fitting function was chosen to match the minimum of the gray curve. The abscissa of this minimum is at the spatial frequencies of the modulation radius, where there is a loss of sensitivity of the PWFS \citep{Guyon_2005}.

The differences between the individually measured optical gains and the fits, which we call the fit residuals, are shown in the bottom plots. In the~0.4" to~1.5" seeing cases, the residuals are below~0.03, which is less than~$5\,\%$ of the optical gain value.  In the~2" case, the residuals of the linear + order~2 and linear + order~3 fit functions are still below~0.04, which is less than~$10\,\%$ of the optical gain value. However, the order~4 fit does not reproduce the shape of the gray curve. Therefore, we chose the linear + order~2 function to fit the modal optical gains. The fit was computed with 12 modes distributed along the basis. More modes yield the same accordance between fit and measurements, whereas with fewer modes, the fit starts to be less reliable (study not shown here).

\subsection{Parametric optimization}

\begin{figure}
    \centering
    \includegraphics[width=0.9\linewidth]{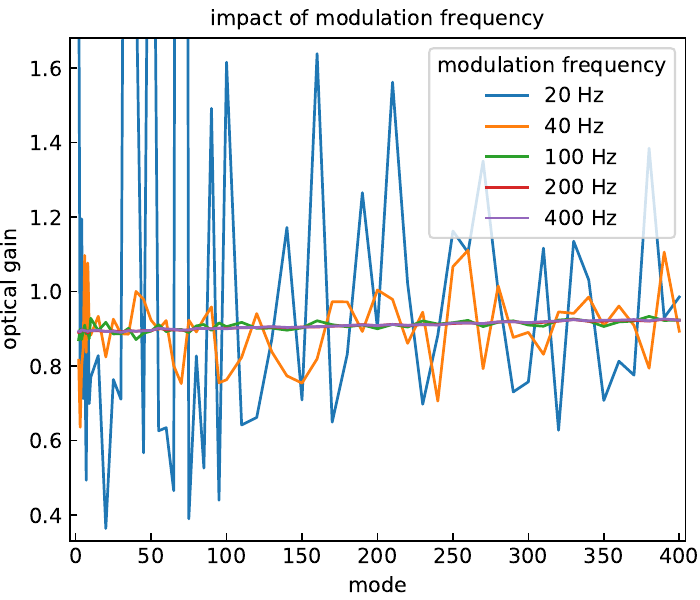}
    \caption{Measured optical gain versus the mode order, for several frequencies from 20 Hz in blue to 400 Hz in purple. Science case: Bright, seeing = 0.8", $\tau_0 = 3$ ms.}
    \label{fig:freq_og}
\end{figure}

The parameters we can tune to optimize the method are the modulation frequency, $f$, the modulation amplitude, $a$, and the exposure time per mode, $t_{modu}$. First, we set the modulation amplitude at $a = 10$ nm RMS and the exposure time at $t_{modu} = 0.5$ s and we explore several values of modulation frequency $f$. We measured the modal optical gains of several modes in the basis, without considering the fit method for now. The optical gains curves versus mode order are presented in Figure~\ref{fig:freq_og} for various modulation frequencies. The shape seen in Figure~\ref{fig:fit_og} (decreasing until mode order~20 and increasing after) is not seen because the scale is too large. The optical gains estimation at a 20 Hz modulation frequency is quite noisy, with variations on the order of 1 in the optical gain value. In this case, the modulation signal is too slow and the AO loop corrects for it. Then, the signal-to-noise ratio of the optical gain measurement increases as the modulation frequency increases. At a 200 Hz modulation frequency and above, the optical gain measurement has a sufficient signal-to-noise ratio to be used. These observations are consistent with the bandwidth of the second-stage AO system, around 300 Hz. Indeed, below this bandwidth, the sinusoidal voltages applied on the mirror are corrected by the second-stage integrator. At a modulation frequency around the AO system bandwidth or higher, the modulation signal on the mirror is too fast to be corrected by the loop. In this case, the demodulation is efficient and the optical gains of the mode can be retrieved. As a precaution, we chose a modulation frequency of $f = 200$ Hz, as far as possible from the AO loop sampling frequency.

\begin{figure}
    \centering
    \includegraphics[width=0.9\linewidth]{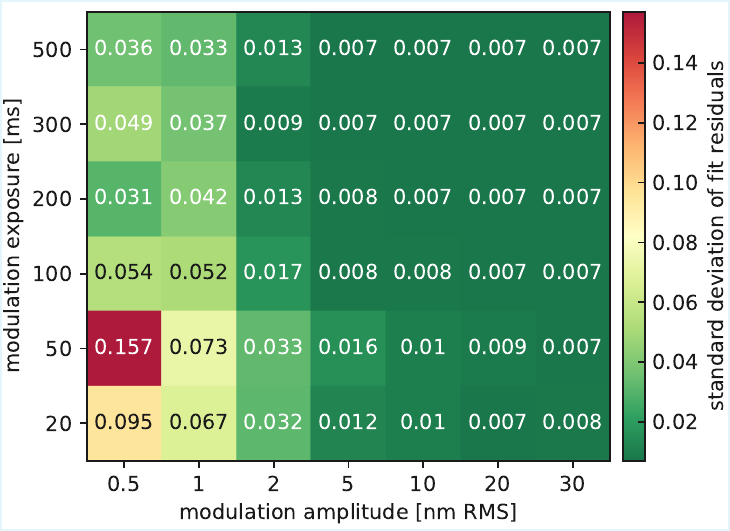}
    \caption{Standard deviation of the fit residuals versus the modulation amplitude and the modulation exposure. Science case: Bright,  $\tau_0 = 2$ ms, seeing = 1".}
    \label{fig:param_og}
\end{figure}

Then we included the fit method described in the previous section. We explored the modulation amplitude, $a$, from 0.5 to $30\,$nm RMS and the exposure time, $t_{modu}$, from 20 to 500 ms. We recall that for one mode, we applied the sinusoidal modulation at the frequency of $f=200$ Hz with a modulation amplitude $a$ for a duration $t_{modu}$. The optical gain of this mode was computed with equation \ref{eq:og}. The method was repeated for the 12 modes along the basis and the linear + order 2 fit was applied. In Figure \ref{fig:param_og} we report the standard deviation of the fit residuals against the modulation amplitude $\vec{a}$ and the exposure time $\vec{t_{modu}}$. We present here the result for a~1" seeing turbulence condition and a~$3\,\lambda_{\mathrm{WFS}} / D$ pyramid modulation radius.
In the upper right part, for high exposure time and modulation amplitude, the standard deviation of the fit residuals is~0.007, less than~$1\,\%$ of the optical gain values. As the exposure time and the modulation amplitude decrease, the standard deviation of fit residuals increases, up to~$0.157$ for high fit residuals. To choose the optimal value of the modulation amplitude, we also have to take into account the limited range of the second stage DM actuator of SAXO+. As the modulation signal will be applied in closed loop, we do not want to hit the stroke limit. Similarly, we would like the optical gain calibration to be as fast as possible and we chose a low exposure time per mode. Considering the results for the other seeing values (0.4", 0.7", 1.5", and 2") and pyramid modulation radius (2 and 1 $\lambda_{\mathrm{WFS}} / D$), not shown in this paper, we conclude that a~$20\,$nm RMS modulation amplitude and a~$0.1\,$s exposure time are a relevant trade-off. In the rest of the paper, when we measure the PWFS optical gains, we use these values and $f = 200\,$Hz as well as the linear + order~2 fit procedure.

In this work, we use the optical gain calibration as is explained in this section, but the method can still be improved. To speed up the calibration, the different modes could be modulated at the same time, with different modulation frequencies. Moreover, the effective modulation amplitude on the DM depends on the modulation frequency, whether this frequency is in the attenuation bandwidth of the transfer function or in the overshoot. It could be relevant to investigate the actual amplitude applied on the DM for various modulation frequencies. These studies are outside of the scope of this paper.

\section{NCPA compensation}
\label{sec:ncpa}

In the rest of the paper, we aim to decrease the stellar speckle intensity in the coronagraph image. In this section, we study the NCPA correction (i.e., offsetting the PWFS reference to minimize the NCPA phase in the pupil upstream from the coronagraph), compensating for the PWFS optical gains or not. When used, the PWFS optical gains are measured as is described in section~\ref{sec:opti_gains}. We do not include the optical gains in the AO loop control law, as we want to quantify their impact on the NCPA correction only. Moreover, we observed that considering the optical gains in the AO loop has no significant impact on the SAXO+ system performance \citep{Goulas_2024}.

As we are only interested in the measured NCPA offset on the AO loop, we assume that we know the NCPA phase perfectly. To do so in numerical simulations, we first opened the AO loops with no turbulence (meaning an ideal AO system lighted by a point-like internal source). We added the NCPA phase screen in front of the PWFS and we recorded the associated PWFS measurement. We set this point as the reference offset of the AO loop so that when closed, the AO loop added the NCPA phase in the coronagraph channel. This way is idealistic and only doable in simulations. In the real system, the NCPA measurement and the calibration of the PWFS offset will be done thanks to a NCPA WFS such as the Zernike sensor.

Then, we assumed that the AO system was observing a given star (different magnitudes) in specific conditions (seeing, coherence time, PWFS modulation, etc). We also compensated for the known NCPA to minimize the phase aberrations in the pupil in front of the coronagraph compensating for the PWFS optical gains (Eq.\,\ref{eq:offset_with_bias}) or not.

\begin{figure}
    \centering
    \includegraphics[width=0.99\linewidth]{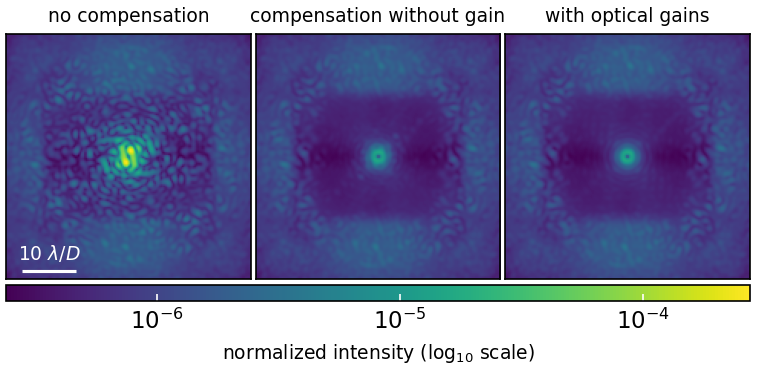}
    \caption{Coronagraph images obtained after a 10 seconds exposure. Left : With NCPA, not compensated for. Middle : Compensation of NCPA without calibrating optical gains. Right : Compensation of NCPA, calibrating optical gains. Seeing = 0.85", $\tau_0 =$ 2 ms.}
    \label{fig:images_ncpa_comp}
\end{figure}

Once the PWFS reference slopes are offset to account for the NCPA, the phase aberrations in the coronagraph channel should decrease and so should the speckle intensity. Figure \ref{fig:images_ncpa_comp} shows the coronagraph images obtained after a 10 second exposure time, with a 0.85" seeing and bright star. In the left image, we did not correct for NCPA and the exoplanet detection capabilities are limited by the speckles. Then, in the middle image we compensated for the NCPA, ignoring the optical gains. We distinguish here the second stage correction zone, which is smaller than the first stage one. Most of the speckles are attenuated below the AO halo, but some of them close to the optical axis are still visible. If we compensate for the PWFS optical gains (right image) the NCPA speckles are more attenuated, especially close to the star position (center of the image).

\begin{figure*}
    \centering
    \includegraphics[width=0.9\linewidth]{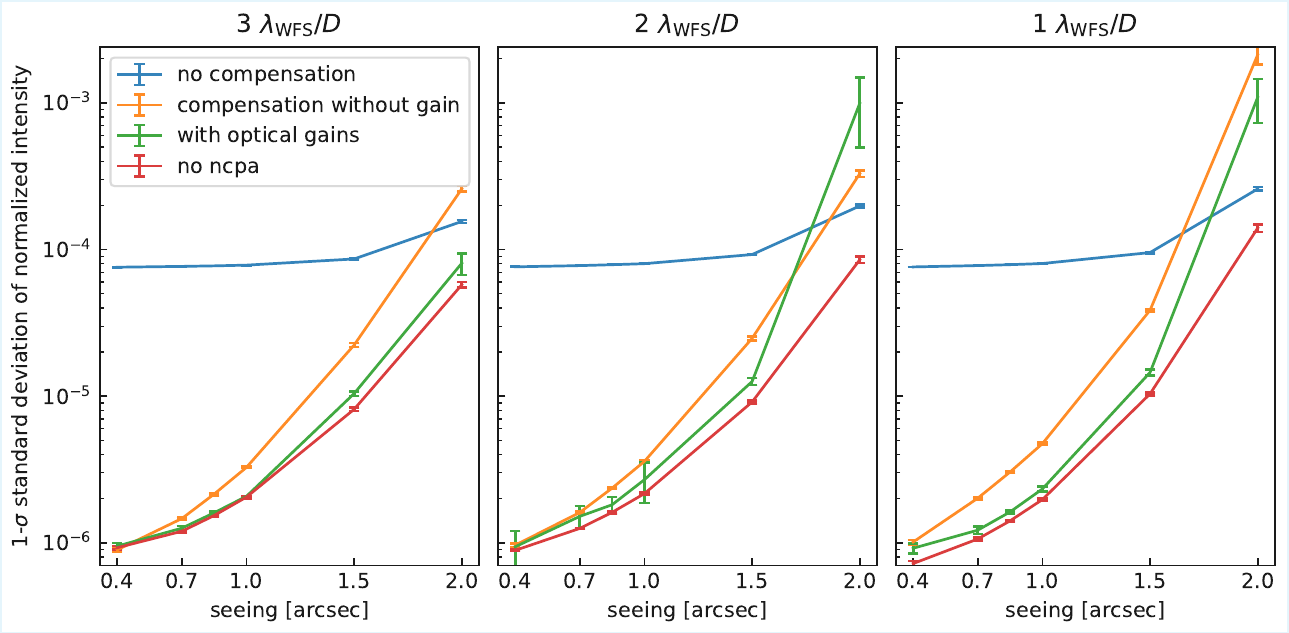}
    \caption{1$\sigma$ standard deviation of coronagraph intensity versus the seeing, for a bright case, $\tau_0 = 2\,$ms, and different pyramid modulation radii : 3, 2, and 1 $\lambda_{\mathrm{WFS}} / D$. We compare the performance of the system without correcting the NCPA (blue), correcting the NCPA with the PWFS reference offset but without optical gains (orange), including an optical gain calibration (green), and the performance without NCPA, which we want to retrieve (red).}
    \label{fig:ncpa_comp}
\end{figure*}

To quantify this improvement, the metric we use is the spatial standard deviation of the normalized intensity. In this section, it was computed with the pixels located in the region between 4 and 6 $\lambda / D$. Figure \ref{fig:ncpa_comp} shows the standard deviation of the coronagraph intensity between 4 and 6 $\lambda / D$ versus the seeing, for three PWFS modulation radius. In blue, there is no NCPA correction. In orange, we show how we compensateds for the NCPA by offsetting the PWFS reference ignoring the PWFS optical gains. Then (shown in green) we applied the previous method to calibrate the optical gains (Section~\ref{sec:opti_gains}) and account for them when offsetting the PWFS reference measurement (Eq.\,\ref{eq:offset_with_bias}). Finally, the red curve is the intensity reached without NCPA in the simulation (limited by AO residuals only), in other words, the performance we would like to retrieve by compensating for NCPA.

NCPA compensation decreases the star speckle intensity by more than a factor of~20 for seeing below~$1"$ for the three PWFS modulations, with a perfect coronagraph. It reaches the performance of the ideal case, where no NCPA are present. The PWFS optical gain compensation is significant for seeing above~0.85", with a gain of a factor of~1.5-2 in speckle intensity, but its impact is negligible for better seeing. In the latter case, the optical gains are close enough to 1 for the NCPA compensation to reach the best performance without estimating the optical gains. At a~2" seeing, the NCPA compensation without optical gains is worse than the no-compensation case. In this case, the PWFS optical gains are significantly below 1 and the NCPA is overcompensated for by the DM.

\begin{figure*}
    \centering
    \includegraphics[width=0.9\linewidth]{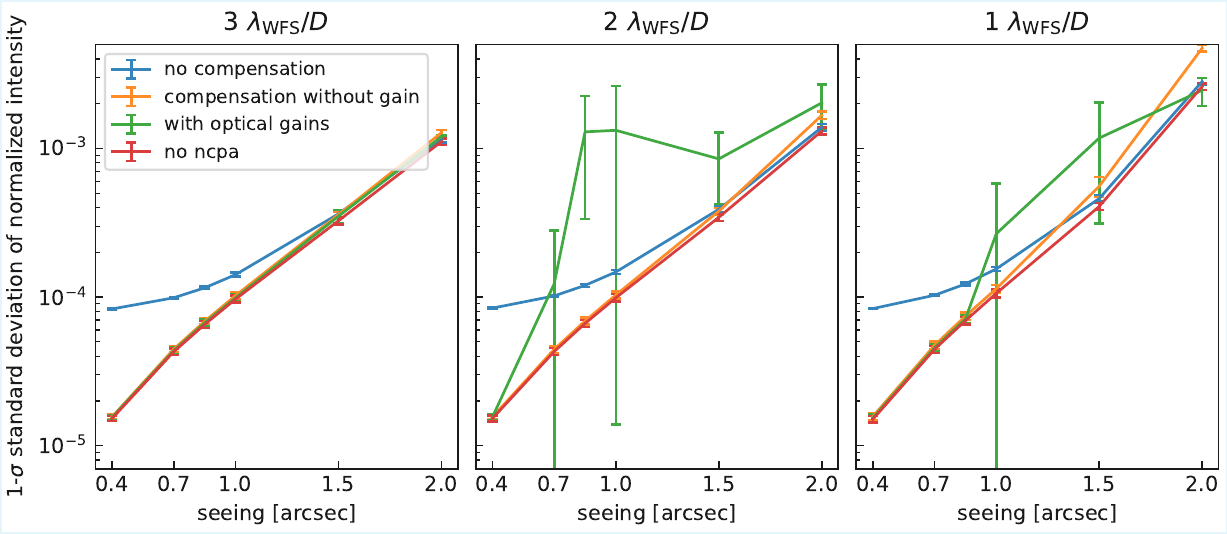}
    \caption{Same as~Fig.\,\ref{fig:ncpa_comp} for a faint case.}
    \label{fig:ncpa_comp_red}
\end{figure*}
Figure \ref{fig:ncpa_comp_red} shows the same plots as Figure \ref{fig:ncpa_comp} for the faint science case. The NCPA compensation ignoring the PWFS optical gains (orange curve) decreases the star speckle intensity down to the no-NCPA case. Accounting for the PWFS optical gains (green curves) does not improve the performance and sometimes results in a bad performance (e.g., $2\,\lambda_{\mathrm{WFS}}/D$ PWFS modulation case). This is because the method presented in section~\ref{sec:opti_gains} is not accurate (noisy estimation of the PWFS optical gains), leading to an incorrect NCPA compensation.

We demonstrate in this section that, for seeings below~1.5", the NCPA compensation enables one to reach the speckle intensity level of a no-NCPA case for both bright and faint targets. We also show that the compensation of the PWFS optical gains is useful to gain a factor of up to~2 in contrast in the coronagraph image for seeing above~1.5".

\section{Dark hole loop}
\label{sec:dh}

The dark hole loop is another method that can be used to attenuate the quasi-static speckles induced by both phase and amplitude aberrations. It aims to minimize the starlight intensity in a given area of the coronagraph image (Section~\ref{sec:ncpa_comp}). The dark hole loop is run during the observation, while the AO loop is closed.

\subsection{Probe with the pyramid wavefront sensor}
\label{subsec:probe}

The PWP technique relies on a temporal modulation of the speckle intensity in the coronagraph image to retrieve the associated electric field~$E$. In this paper, we consider the recording of four successive probe images, each of them being the long exposure coronagraph image recorded while the AO loop is closed and one actuator of the DM (adding a probe phase) is pushed or pulled. Assuming a linear relation between the probe phase and its influence in the coronagraph image, we multiply the four probe images by a pair-wise matrix to retrieve~$E$. On-sky, the pair-wise matrix has to be estimated with a synthetic model of the instrument. In our simulations, we use the instrument simulator to calibrate this pair-wise matrix.

To apply a probe on the DM while the AO loop is closed, we offset the PWFS reference. As for NCPA compensation, due to the PWFS nonlinearities, this method could yield a distorted probe, meaning the probe phase effectively applied on the~DM is not the one assumed in the synthetic model. Therefore, the PWP and EFC optical model will be mistaken, inducing errors in the estimated~$E$ field and on the DM shape used to minimize its intensity inside the dark hole.

We first simulated a single-stage AO system, at 3\,kHz with a PWFS, in a bright star scenario. The configuration is the same as the one described in Section~\ref{sec:sim_assumptions} without the first stage. The coherence time is $\tau_0 = 3\,$ms and the seeing = 1.1".

\begin{figure}
    \centering
    \includegraphics[width=0.9\linewidth]{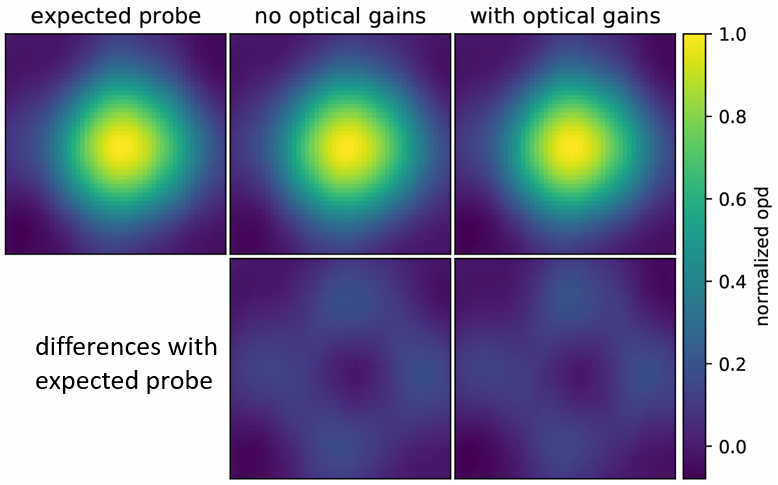}
    \caption{Pupil plane images zoomed in around the shape of the probe applied on the DM. The optical path difference is normalized to the maximum of the probe. Top row, left : Expected probe, namely the influence function of the actuator. Middle : Shape of the probe in closed loop without considering optical gains. Right : Same as middle but including an optical gain calibration. Bottom row : Differences between the obtain probe in closed loop and the expected probe. Science case : Bright-1, seeing = 1.1", $\tau_0$ = 3 ms.}
    \label{fig:probe_shape}
\end{figure}

In this paragraph, we assume that the probe image exposure time is 10\,s. While the AO loop was closed, we applied a 300\,nm probe on the DM changing the PWFS reference offset (Section~\ref{eq:offset_with_bias}). We then measured the phase that was effectively applied to the system, averaging~10,000 residual phase screens during the closed loop run. The residual turbulent phase averages to zero, bringing out the static phase (i.e., the probe phase). In the first row of Figure~\ref{fig:probe_shape}, the left image is the shape of the probe that we expect to apply to the mirror. In other words, this is the influence function of the actuator we poke (that we have access to in simulation). The second and third columns show the shapes of the probe obtained by offsetting the PWFS reference, without and with optical gain compensation respectively. The second row shows the differences between the expected probe and its actual shape.
The optical path difference is normalized to the maximum of the probe in each image so that wa can see the impact of the PWFS nonlinearities on the shape of the probe and not the amplitude. There is no clear difference between the expected shape and the one we get while the AO loop is closed, with or without PWFS optical gain calibration. The image differences (second row) are always better than~20\,\% of the maximum of the amplitude of the probe. We conclude that the PWFS optical gains have no significant impact on the shape of the probe. In \citet{Potier_2022}, the shape of the influence function of the SAXO DM is known with a similar accuracy and it does not limit the dark hole correction.

\begin{figure}
    \centering
    \includegraphics[width=0.9\linewidth]{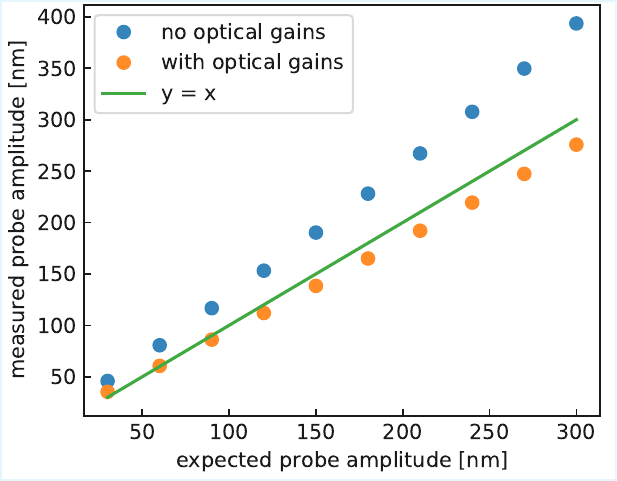}
    \caption{Obtained amplitude of the probe in closed loop versus the expected amplitude, without considering optical gains in blue and including the optical gain calibration in orange. Science case : Bright-1, seeing = 1.1", $\tau_0$ = 3 ms.}
    \label{fig:probe_amp}
\end{figure}

The amplitude of the probe is also an important parameter that has an impact on the convergence time of the dark hole loop. We repeated the previous simulation with probe amplitudes ranging from~30 to~300\,nm. Figure~\ref{fig:probe_amp} shows the measured probe amplitude versus the expected one. The blue dots are the probes ignoring the PWFS optical gains and the orange ones the probes  accounting for them. The $y = x$ curve (best scenario) is shown in green.
The effective amplitude is closer to the expected one when PWFS optical gains are used. For a probe expected to be 300\,nm, we obtain a~400\,nm probe, ignoring the optical gains. This overshoot is expected as the PWFS optical gains are lower than~1. After calibration of the PWFS optical gains, the probe is at~270\,nm, corresponding to a~10\,\% error in the probe amplitude. Therefore, for certain observation conditions, it may be necessary to use the PWFS optical gains to have a better knowledge of the amplitude of the probe that is used for PWP. We remind the reader that because the PWP is model-based, the knowledge of the probe phase is directly linked to the accuracy of the speckle electric field that is measured and to the speed of the dark hole loop convergence.

\subsection{Dark hole loop along a single-stage AO with a PWFS}
\label{subsec:DH_1PWFS}
In this section, we study the dark hole loop in cascade of a single-stage AO system running at~3\,kHz with a PWFS. A parametric study of the dark hole loop along the SAXO+ two-stage AO is the subject of the next section. The observing conditions are the bright target scenario with a 1" seeing and a coherence time of~$\tau_0 =2\,$ms. Four iterations of the dark hole loop are run, with a dark hole gain of~0.8. The simulated coronagraph is the SPHERE APLC, but we do not simulate either the photon or the readout noise in the coronagraph image (not yet implemented in the COMPASS tool). The absolute performance is optimistic, but we therefore emphasize the impact of the PWFS optical gains on the dark hole loop.

During one iteration of the dark hole loop, we sensed the  electric field of the speckles in the coronagraph image and we modified the DM shape to minimize their intensity. The sensing was done by PWP using two side-by-side actuators as probes \citep{Potier_2022}. Each actuator was successively pushed then pulled while the AO loop was closed, by applying an offset on the PWFS reference. Therefore, we recorded four images of~45\,s exposure each. Using the PWP synthetic model, we retrieved an estimation of the static focal plane electric field over 45s-AO residuals, which was enough to average the turbulence \citep{singh2019}. With this estimation and a synthetic interaction matrix between the DM modes and the coronagraph electric field, the EFC method yielded a DM shape that minimizes the speckle intensity in a given area of the coronagraph image called the dark hole. We applied this DM shape in closed AO loop, again offsetting the PWFS reference, and recorded a 45\,s coronagraph image. After a few iterations, we aim to obtain coronagraph images showing a dark hole with almost no speckles inside. Here, we assume that the dark hole is a half ring between~3 and $9\,\lambda / D$ centered on the optical axis. We chose a half-dark hole to minimize the stellar light induced by phase aberrations (NCPAs) and pupil diffraction (spiders and imperfections of the coronagraph).
\begin{figure}
    \centering
    \includegraphics[width=0.99\linewidth]{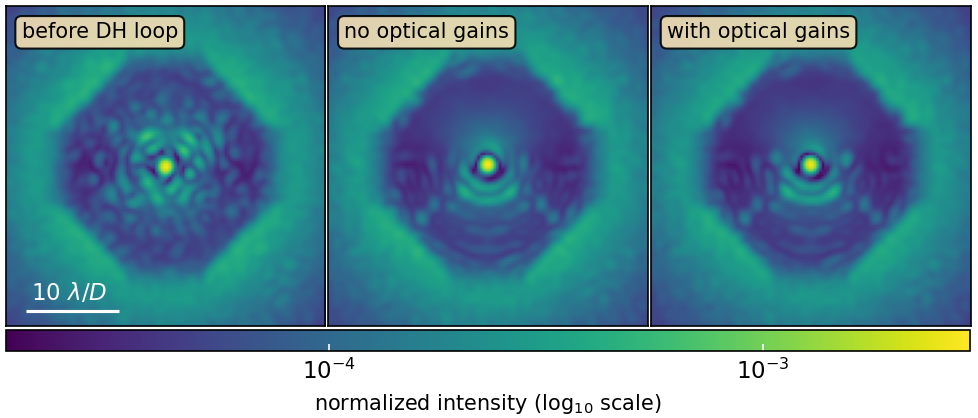}
    \caption{Coronagraph images before the dark hole loop (left) and after four dark hole iterations, without the PWFS optical gains (middle) and with the PWFS optical gains (right). }
    \label{fig:coro_images_pyr_stage}
\end{figure}

Figure \ref{fig:coro_images_pyr_stage} compares the coronagraph images obtained before the dark hole loop (left) and after four iterations (middle and right). The middle image was obtained ignoring the PWFS optical gains and the right image accounting for them. In that case, we ran the PWFS optical gain calibration before each iteration of the dark hole loop. Such a calibration that lasts 2\,s (cf. Section~\ref{sec:opti_gains}) can be done during the reading of the coronagraph image detector. The PWFS optical gain compensation was applied both for the PWP sensing and for the EFC correction. In the left image, the AO loop correction zone extends to 10\,$\lambda / D$ from the optical axis. This limit is set by the number of modes controlled in the AO loop. Inside this area, the coronagraph image is speckle-dominated (NCPAs and static diffraction pattern). After the dark hole loop (middle and right), the static aberrations due to the NCPAs and the coronagraph diffraction are minimized inside the dark hole area. However, it is hard to measure how much the DH reduces the static speckle intensity by because the AO halo dominates the intensity.

\begin{figure}
    \centering
    \includegraphics[width=0.99\linewidth]{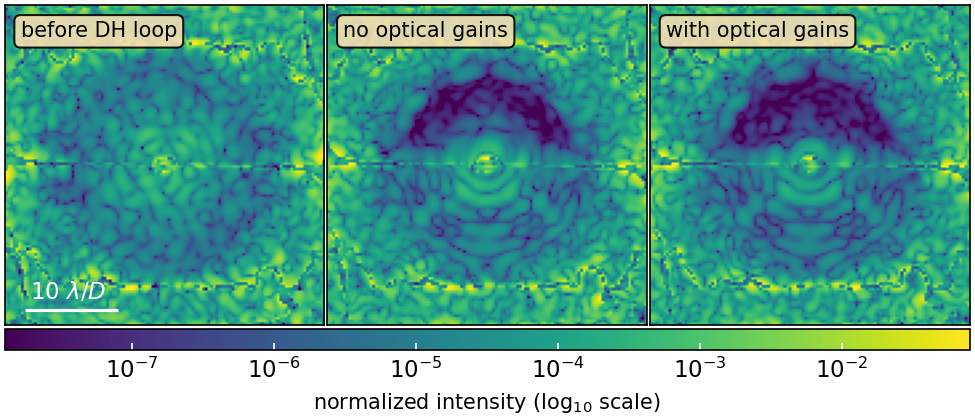}
    \caption{Coherent intensity images, before the dark hole loop (left) and after four dark hole iterations, without the PWFS optical gains (middle) and with the PWFS optical gains (right).}
    \label{fig:pw_int_pyr_stage}
\end{figure}
To measure the reduction of the static speckle intensity, we used the intensity estimated by the PWP method. This estimate only probes the speckle intensity (static during the 45\,s exposure). Usually known as the coherent intensity, it does not measure the intensity of the AO halo. The intensity in the raw coronagraph image is called the total intensity. Figure~\ref{fig:pw_int_pyr_stage} shows the coherent intensity for the three cases represented in Figure~\ref{fig:coro_images_pyr_stage}, and hence shows only the static speckles. At the DM cutoff frequency of~13\,$\lambda / D$ and further, the PWP method cannot estimate the static electric field, and therefore the measure is not reliable. Also, as we used two probe actuators, the speckles are not well estimated on the horizontal central line~\citep{Potier_2022}. Before the dark hole loop (left image), the image shows the NCPA speckle pattern introduced in the simulation. After four iterations of the dark hole loop, the performance is clearly improved if we compensate for PWFS optical gains (right) over the case in which we ignore them (center), especially for the speckles close to the optical axis.
\begin{figure}
    \centering
    \includegraphics[width=0.9\linewidth]{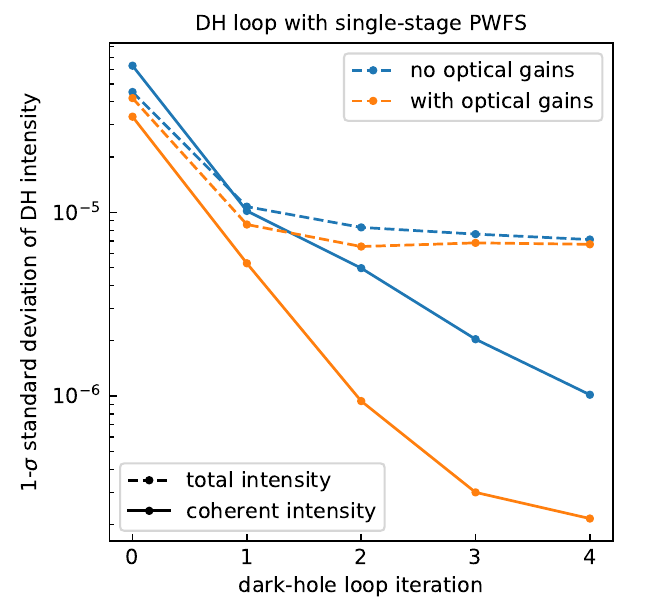}
    \caption{Single-stage AO PWFS loop and dark hole loop. Standard deviation of the total intensity (dashed line) and the coherent intensity (solid line) in the DH region versus iteration number of the dark hole loop. Science case : Bright, seeing = 1", $\tau_0 =$ 2 ms.}
    \label{fig:coro_vs_pw_int}
\end{figure}

To quantify this improvement, the metric we use is the spatial standard deviation of the coherent intensity. In this section, it is computed with the pixels located in the DH region. Figure \ref{fig:coro_vs_pw_int} shows the 1$\sigma$ standard deviation inside the dark hole area for both intensities (from the Figure~\ref{fig:coro_images_pyr_stage} coronagraph images in dashed lines or the Figure~\ref{fig:pw_int_pyr_stage} speckle images in solid line) versus the number of iterations of the dark hole loop. The blue curves correspond to the dark hole loop ignoring the PWFS optical gains, and the orange curves correspond to the dark hole loop accounting for the PWFS optical gains.

At iteration 0, before the dark hole loop starts, the standard deviation of the coronagraph intensity starts at $\sim5 \cdot 10^{-5}$. The coherent intensity is higher when the PWFS optical gains are ignored. As these gains are below 1, the amplitude of the probe applied during the PWP is higher than expected by the model (section~\ref{subsec:probe}). Consequently, ignoring the PWFS optical gains, the PWP algorithm overestimates the static intensity.

For the total intensity, the standard deviation decreases by a factor of $<10$ in two iterations and then remains constant, whether or not we calibrated the PWFS optical gains. This level is set by the residual AO-halo that is asymmetric because of wind. As one looks for point-like sources (e.g. exoplanets), the coherent intensity is a better criterion to probe the performance of the dark hole technique. The standard deviation of the coherent intensity decreases during the four iterations by a factor of 70 when ignoring the PWFS optical gains and by a factor of over 100 when compensating for the PWFS optical gains.

This result shows the benefit of calibrating and compensating the PWFS optical gain during a dark hole loop behind a single pyramid AO loop.

\subsection{Dark hole loop with the two-stage SAXO+ system}

\begin{figure}
    \centering
    \includegraphics[width=0.9\linewidth]{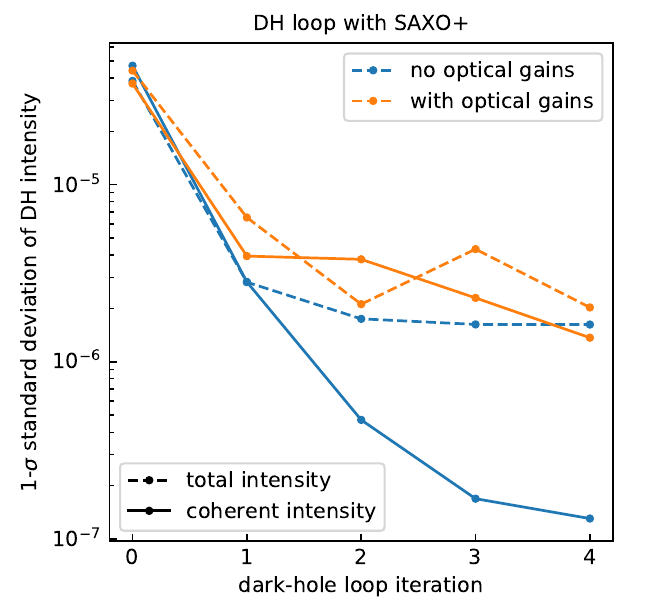}
    \caption{Two-stage AO (SAXO+ system) and dark hole loop. The standard deviation of intensity is plotted versus the iteration number of the dark hole loop. The dashed line shows the standard deviation of the intensity in the total coronagraph image, while the solid line corresponds to the coherent intensity. Science case : Bright, seeing = 1", $\tau_0 =$ 2 ms.}
    \label{fig:coro_vs_pw_int_2stages}
\end{figure}

In this section, we study the three-loop system composed of the two-stage AO of SAXO+ followed by the dark hole loop. To quantify the impact of the PWFS optical gains, we ran the same simulations as for section~\ref{subsec:DH_1PWFS} replacing the AO loop by the two AO loops described in Tab~\ref{tab:param_1}. Figure~\ref{fig:coro_vs_pw_int_2stages} shows the 1$\sigma$ standard deviation inside the dark hole area of the total intensity (dashed lines) and the coherent intensity (solid lines) versus the number of iterations of the dark hole loop. In blue, the PWFS optical gains are ignored, and in orange, they are compensated for.

Before the dark hole loop (iteration 0) the standard deviation of intensity starts at $4 \cdot 10^{-5}$. We note that even when ignoring the PWFS optical gains, the coherent intensity (solid line) is at the correct value. This is because the first stage of the AO loop produces a PSF on the PWFS that is good enough for the PWFS optical gain to be close to~1. Hence, the coherent intensity estimate is not biased as it was in~Figure~\ref{fig:coro_vs_pw_int}.

The standard deviation of the total intensity (dashed lines) decreases by a factor of~20, which is an improvement with respect to the single-stage pyramid AO case of Figure~\ref{fig:coro_vs_pw_int}. The two-stage SAXO+ AO loop reduces the AO residual halo by a factor of 4.

When we consider the coherent intensity, the standard deviation decreases by a factor of over~100 if the PWFS optical gains are ignored (solid blue line). However, the performance degrades if we account for these gains (dashed blue line). We noticed that the calibration method of section~\ref{sec:opti_gains} is not robust at a high Strehl ratio, and provides very noisy measurements of the optical gain. The PWP algorithm therefore yields a bad estimation of the static electric field, and the EFC correction is not accurate. In section \ref{sec:ncpa}, the optical gains calibration was successful because NCPA compensation is less impacted by errors on optical gains than the PWP + EFC loop. In section \ref{sec:ncpa} we assumed that we knew the NCPA phase screen, while in section \ref{sec:dh} we estimated the static speckles with PWP. The measurement errors on the optical gains propagate on the estimation of the static electric field in the focal plane. In the rest of this section, we assume the PWFS optical gains are no longer compensated.

\begin{figure}
    \centering
    \includegraphics[width=0.99\linewidth]{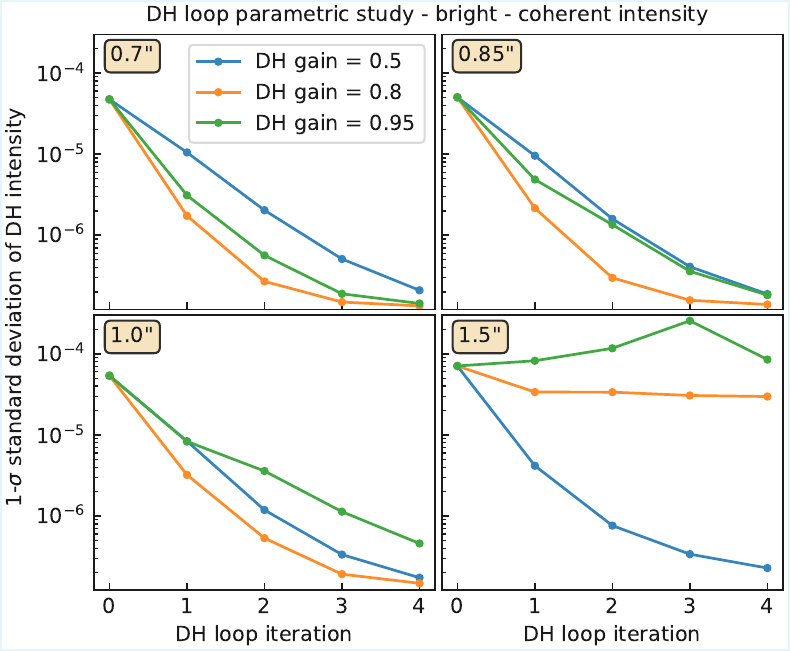}
    \caption{Parametric study of the gain of the dark hole loop for a bright case with $tau_0 = 2\,ms$. . Each subplot shows the standard devation of the intensity estimated by PWP (inside the dark hole area), versus the iteration number of the dark hole loop.}
    \label{fig:dh_parametric_study_bright}
\end{figure}

\begin{figure}
    \centering
    \includegraphics[width=0.99\linewidth]{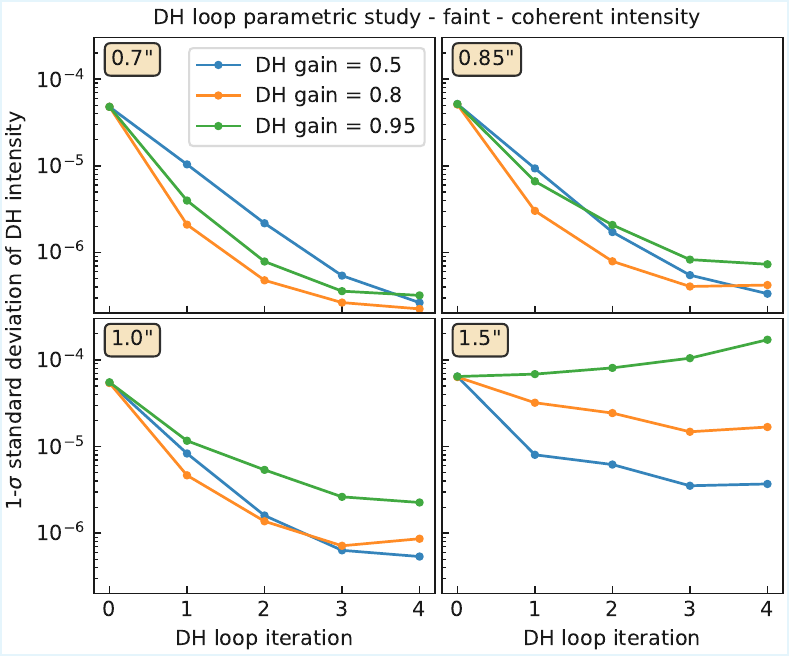}
    \caption{Parametric study of the gain of the dark hole loop for a faint case with $tau_0 =2\,$ms. . Each subplot shows the standard devation of the intensity estimated by PWP (inside the dark hole area), versus the iteration number of the dark hole loop.}
    \label{fig:dh_param_study_red}
\end{figure}

The gain of the dark hole loop is an important parameter that balances the trade-off between the speed of convergence and the stability of the dark hole loop. Figure \ref{fig:dh_parametric_study_bright} shows a parametric study of the dark hole loop for a bright star case. It plots the $1\,\sigma$ standard deviation of the coherent intensity against the iteration number of the dark hole loop for several dark hole loop gains: 0.5 in blue, 0.8 in orange, and 0.95 in green. Each subplot corresponds to one seeing (0.7", 0.85", 1.0" and 1.5"). Figure \ref{fig:dh_param_study_red} is the same figure for a faint star case.

For the seeings below~1" and both bright or faint stars, the~0.8 dark hole gain leads to the best performance: the final performance is reached in four iterations. For the bright star, the speckle intensity is reduced by a factor of~500, reaching a~$10^{-7}$ contrast level. For the faint star, the reduction factor goes from 500 for a 0.7" seeing to 100 for a~1" seeing. For the faint star, we also notice that the 0.5 dark hole gain leads to a similar and sometimes better performance (0.85" and 1" seeing). For a 1.5" seeing, the 0.5 dark hole gain leads to a better performance in both star cases because it reduces the impact of the AO-residual halo on the PWP probe images. And even with a~1.5" seeing, the dark hole reduces the static speckle intensity by a factor of~500 for the bright star and~10 for the faint star.

\begin{figure}
    \centering
    \includegraphics[width=0.99\linewidth]{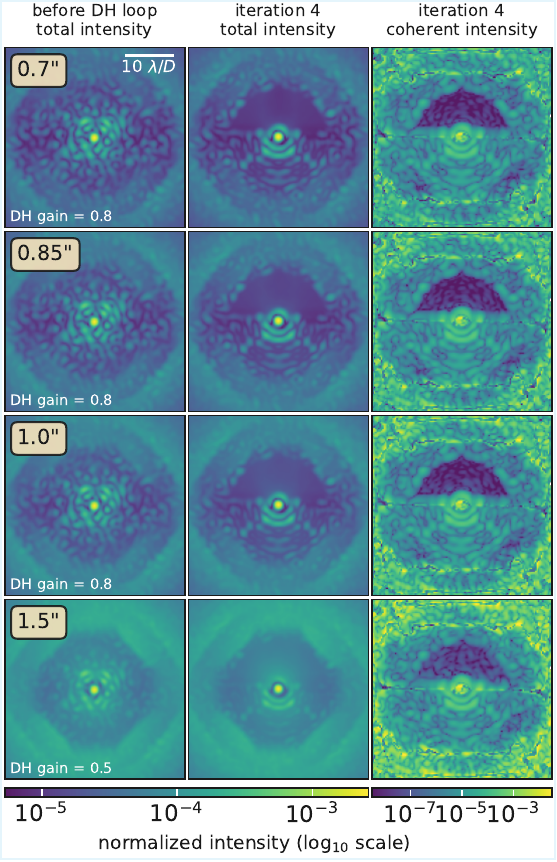}
    \caption{Total intensity and coherent intensity images of the dark hole parametric study for the bright case.}
    \label{fig:im_2AO_dh_bright}
\end{figure}

\begin{figure}
    \centering
    \includegraphics[width=0.99\linewidth]{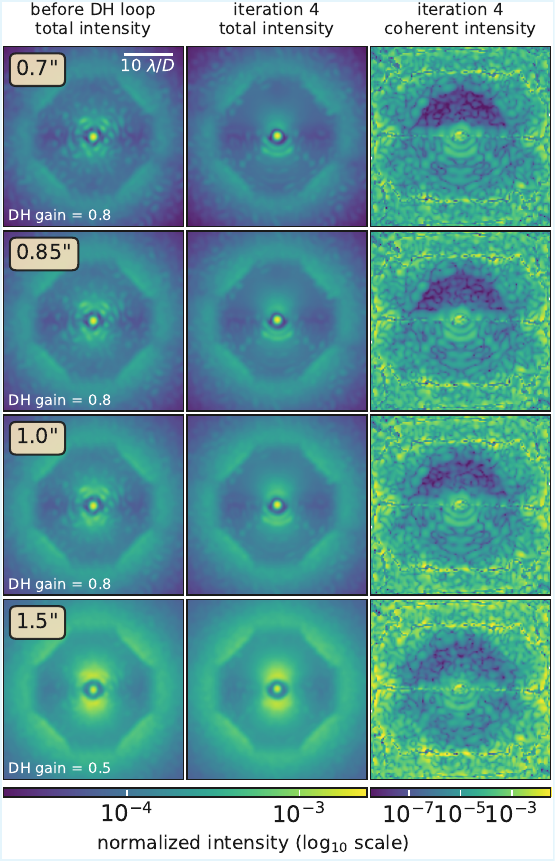}
    \caption{Total intensity and coherent intensity images of the dark hole parametric study for the faint case.}
    \label{fig:im_2AO_dh_red}
\end{figure}

Figure \ref{fig:im_2AO_dh_bright} and \ref{fig:im_2AO_dh_red} show the coronagraph images before the dark hole loop and at iteration 4 (first column and second column), and the coherent intensity at iteration 4 (third column). Figure \ref{fig:im_2AO_dh_bright} is for the bright science case and Figure \ref{fig:im_2AO_dh_red} is for the faint science case. Each row corresponds to one simulated seeing. The images are presented with the optimal dark hole loop gain, which is 0.8 for the 0.7", 0.85", and 1" seeing case and 0.5 for 1.5" seeing. In the first column, the speckles from the NCPAs are brighter than the AO residual halo. Then, in the second column, the dark hole loop reduces the intensity of the speckles below the intensity of the halo. The coherent intensity is useful to measure the residual speckle intensity after the dark hole loop (third column).

Finally, we find no need to compensate for the PWFS optical gains to perform the dark hole technique behind the two AO loops of the SAXO+ system. We also demonstrate that the dark hole technique can reduce the speckle intensity in the coronagraphic image at any seeing up to~$1.5"$ for bright and faint stars. And the reduction factor is between 10 in the worst case (faint star with 1.5" seeing) and 500 in the best case (bright star with 0.7" seeing). 

\section{Conclusions}

SAXO+ is the upgrade of SAXO, the~AO system of the exoplanet imager SPHERE/VLT. It will include a second-stage AO, downstream of the SAXO stage, with a near-infrared PWFS. Because of the NCPAs, the exoplanet detection capabilities of the instrument will be limited by the speckle intensity in the coronagraph image. We investigated two strategies to decrease the speckle intensity in the coronagraph images. The first one is a NCPA compensation that minimizes the NCPA phase in the pupil plane. The second one is the dark hole loop that minimizes the speckle intensity in the coronagraph image. The dark hole loop is then a third loop after the two AO loops. Both NCPA compensation and dark hole loop techniques require one to modify the static component of the DM shape, during the closed-loop regime. In both cases, this is done by offsetting the reference of the PWFS measurements.

As the PWFS is quite nonlinear, an offset in its reference may result in the AO loop having an unexpected bias. The sensitivity loss of the PWFS due to its nonlinearities can be described with modal coefficients called optical gains. We adapted a calibration method of the optical gains developed in \citet{Esposito_2020} to SAXO+. It consists in temporally modulating a mode on the DM in closed loop and retrieving the associated optical gain by demodulating the PWFS measurements and the DM commands. Our strategy relies on applying this calibration on~12 modes of the basis and fitting the measured optical gains to estimate the gains over the whole basis. We optimized the time required for each mode measurement and the function that describes the optical gain as a function of the modes. We also optimized the frequency and the amplitude of the modulation. We finally chose a 20\,nm modulation amplitude, 200\,Hz frequency modulation, and a total time of~2\,s for the calibration of the whole basis.

We then presented the impact of the PWFS optical gain compensation on both NCPA compensation and dark hole correction. We demonstrated in section~\ref{sec:ncpa_comp} that, for seeings below~1.5", the NCPA compensation reaches the contrast level of the ideal case in which no NCPA are present, in both the bright and faint cases, and for several PWFS modulation radii. We also show that the compensation of the PWFS optical gains is useful to gain a factor of up to~2 in contrast in the coronagraph image for seeings above~1.5" but has a negligible impact for better seeings.

In section~\ref{sec:dh}, we show that the PWFS optical gains are useful to enhance the dark hole performance behind a single-stage pyramid AO loop. On the contrary, they are not needed if the dark hole loop is behind the SAXO+ two AO loops. In the SAXO+ case, the first AO loop provides a PSF with a high Strehl ratio and the PWFS optical gains are already close to~1. We also demonstrated that the dark hole technique behind the SAXO+ system can reduce the speckle intensity in the coronagraphic image at any seeing up to~$1.5"$ for bright and faint stars. The reduction factors are between 10 in the worst case (faint star with 1.5" seeing) and 500 in the best case (bright star with 0.7" seeing). It is therefore a very attractive technique to probe faint exoplanets with high-contrast imaging instruments. However, these results are optimistic considering that the coronagraph images are noiseless in our simulations. The exposure time of the probes is a parameter to be optimized depending on the star magnitude.

The study presented in this paper will support the implementation of NCPA compensation and dark hole loop algorithms in the SAXO+ upgrade. As a technological demonstrator of a two-loop AO system with a PWFS, SAXO+ is part of the ESO roadmap for the Planetary Camera Spectrograph \citep{Kasper_2021} at the Extremely Large Telescope.

\begin{acknowledgements}
The authors are grateful to the PEPR Origins (ANR-22-EXOR-0002, ANR-22-EXOR-0003, ANR-22-EXOR-0017) for its financial support within the Plan France 2030 of the French government operated by the National Research Agency (ANR).
This work was supported by CNRS/INSU through its commission spécialisée astronomie-astrophysique (CSAA).
\end{acknowledgements}

%
%

\bibliographystyle{aa} 
\bibliography{biblio} 

\end{document}